\documentclass{appolb}
\usepackage{graphicx}

\usepackage{amsmath}
\usepackage[style=phys,giveninits=false,biblabel=brackets]{biblatex}
\usepackage{physics}

\newcommand{\deltaphi}{\Delta\varphi}
\newcommand{\GeVc}{GeV/$c$}
\newcommand{\ptiso}{p_\text{T}^\text{iso, ch}}
\newcommand{\ptratio}{p_\text{T}^\text{jet}/p_\text{T}^\gamma}
\newcommand{\pt}{p_\text{T}}
\newcommand{\snn}{\sqrt{s_\text{NN}}}
\newcommand{\ssvar}{\sigma^2_\text{long(5x5)}}

\bibliography{bib}
\begin{document}
\title{Isolated photon-jet correlations in Pb-Pb collisions at $\sqrt{s_\text{NN}} = 5.02$ TeV in ALICE%
\thanks{Presented at Quark Matter 2022}%
}
\author{Alwina Liu
\address{University of California, Berkeley}
\\[3mm]
{On behalf of the ALICE Collaboration
}
}
\maketitle
\begin{abstract}
Jets correlated with isolated photons are a promising channel to study jet quenching in heavy-ion collisions, as photons do not interact strongly and therefore constrain the $Q^2$ of the initial hard scattering.
We present the isolated photon-jet correlations measured in Pb--Pb collisions at $\sqrt{s_{\text{NN}}}$ = 5.02 TeV by the ALICE collaboration.
We study correlations of isolated photons above 20 GeV/$c$ with charged-particle jets above 10 GeV/$c$, reconstructed with the anti-$k_\text{T}$ algorithm. The correlations probe the lowest jet $p_\text{T}$ range ever measured at LHC energies, and larger modifications due to the QGP are expected in the lower $p_\text{T}$ regime.

\end{abstract}
  
\section{Introduction}
In an ultra-relativistic collision between two nuclei, occasionally two partons will interact to produce particles with large transverse momentum. Sometimes, a hard scattering will produce a photon, known as a prompt photon, which can then traverse the quark-gluon plasma (QGP) produced by the rest of the collision without strongly interacting with the QGP. Therefore, when a jet is produced along with a prompt photon in a hard scattering, while the jet loses energy to the QGP, the photon tags the initial energy of the scattered parton that produces the jet. By studying these photon-jet events, one can study partonic energy loss in the QGP, probe bound nucleon structure, and do detailed studies of jet fragmentation and the potential modification thereof by the QGP. Photon-jet correlations are not a new concept; they have been measured in $\snn=5.02$ TeV Pb-Pb collisions at the LHC by both CMS~\cite{gammajet-cms} and ATLAS~\cite{gammajet-atlas}. However, this measurement considers lower-$\pt$ photons (down to 20 \GeVc) and charged-particle jets (down to 10 \GeVc), which probes a low-$x$, low-$Q^2$ regime.

The data for this measurement was collected in 2018 with the ALICE detector~\cite{alice-experiment,alice-performance} at the LHC from Pb-Pb collisions at $\snn=5.02$ TeV. The Inner Tracking System and Time Projection Chamber are used together to reconstruct charged particles (tracks). The VZERO system is used for the minimum-bias trigger and to determine the centrality of the collisions. The Electromagnetic Calorimeter system (EMCal) is used to measure the photons by grouping neighboring cells into clusters; it is also used to trigger on collisions with a large deposit of energy in the EMCal.

\section{Analysis}
The goal of this analysis is to measure the correlation between isolated prompt photons and the charged-particle jets produced in the same hard scattering. We make an isolation cut on the photon of $\ptiso < 1.5$ GeV/$c$ within $R=0.2$. To reconstruct the jets from tracks with $\pt > 0.15$ \GeVc, we use FastJet~\cite{fastjetmanual} with the anti-$k_\text{T}$ algorithm with $R=0.2$. The angular correlations ($\deltaphi$) and momentum balance ($\ptratio$) of photon-jet pairs are measured for photons with $20 < \pt^\gamma < 40$ GeV/$c$ within $\abs{\eta} < 0.67$ and for charged-particle jets with $\pt^\text{ch jet} > 10$ GeV/$c$ within $\abs{\eta} < 0.7$.

At leading order, prompt photons are produced in isolation, whereas neutral mesons (which decay into photons) and fragmentation photons tend to be produced within jets. The isolation energy of a photon candidate can therefore be used to reduce the contribution of these non-prompt photons. It is defined as the sum of the $\pt$ of the charged tracks within $R=0.2$ of the EMCal cluster after subtracting the underlying event density $\rho$:

\begin{equation}
    \pt^\text{iso, ch} = \Sigma_\text{tracks} \pt^\text{track} - \rho(\pi R^2)
    \label{eq:ptiso}
\end{equation}

The underlying event from the Pb-Pb collisions causes both the reconstructed jet $\pt$ and the photon $\ptiso$ to be too high. The underlying event density $\rho$ is estimated with the jet-area/median method~\cite{jetareamedian} and used to correct both the photon isolation (Eq.~\ref{eq:ptiso}) and the jet $\pt$.

The underlying event also causes ``fake'' jets to be reconstructed from particles not associated with any hard scattering. Along with these ``fake'' jets, there are also jets produced from a different process than the one that produced the photon; collectively, we call this the combinatorial background. To estimate its size, we use an event mixing technique in which photons from an EMCal-triggered event are paired with charged-particle jets from a different minimum-bias event. Events are matched to each other based on centrality, $z$-vertex position, and event-plane angle. By construction, the photon and jets are not produced in the same hard scattering, so this mixed-event correlation can be subtracted from the same-event correlation to remove the combinatorial background.

The other large source of background is photons from neutral meson decays, and we use the shower profile in the EMCal to distinguish between signal and background photons and measure the purity with a template fit. As in~\cite{gammahadron}, the shower profile of a photon in the EMCal, i.e. the shower shape, is encoded geometrically and used to define the signal region (SR), which contains most of the prompt photons, and the background region (BR), which is dominated by decay photons. Unlike in~\cite{gammahadron}, a variation of the standard shower shape variable is used; it considers the cells surrounding the cell with the most energy in the cluster and is called $\ssvar$.

\begin{figure}[htb]
\centerline{%
\includegraphics[width=12.5cm]{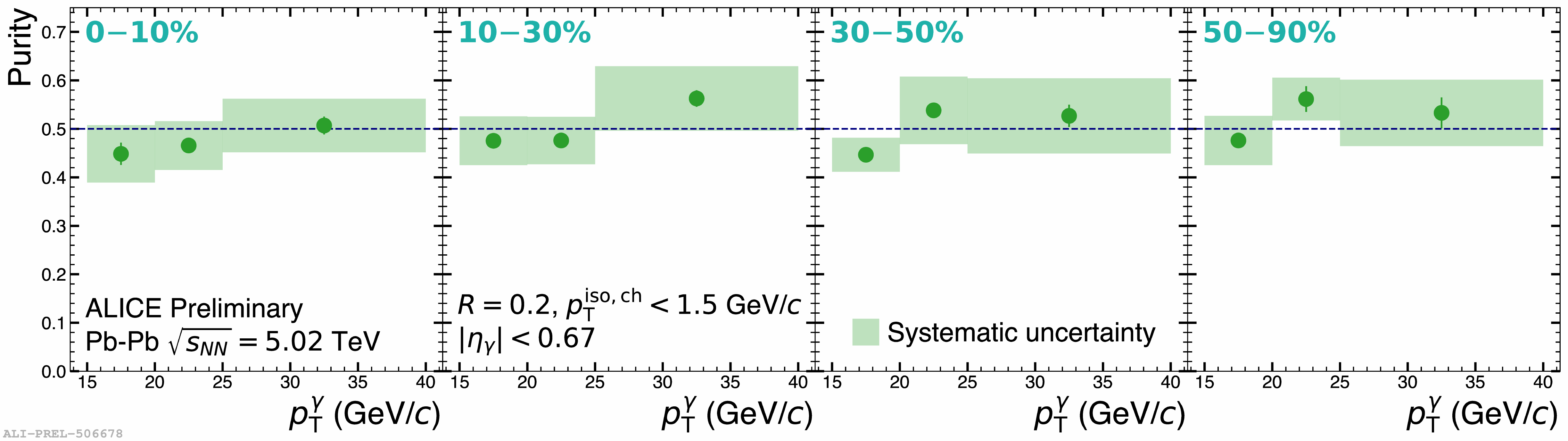}}
\caption{The photon purity is measured as a function of photon candidate $\pt$ across various centrality ranges for photon candidates with $\ptiso < 1.5$ \GeVc.}
\label{fig:purity}
\end{figure}

The prompt photon candidates are selected to be the EMCal clusters that pass the isolation cut and have a signal-like shower shape ($0.1 < \ssvar < 0.3$). However, in addition to the isolated prompt photons we wish to measure, this sample also contains isolated decay and fragmentation photons with a signal-like shower shape. In order to estimate the purity of the photon candidate sample, we fit the shower shape distribution of isolated clusters to a linear combination of templates. This purity is then extracted in bins of centrality and photon candidate $\pt$, and does not change very much with either, as seen in \figurename~\ref{fig:purity}.

With the photon candidate purity, we can estimate and subtract the non-prompt photon background from the photon candidate sample. First, we pair each prompt photon candidate with all charged-particle jets in the event with $\pt > 10$ \GeVc, leading to a per-trigger (i.e. per-photon-candidate) yield of photon-jet pairs, which we call $C_\text{SR}$. This is a linear combination of the per-trigger yield for prompt photons and non-prompt photons, with the relative weight given by the purity. The non-prompt photon correlation can be estimated by the clusters in the shower shape background region ($0.6 < \ssvar < 1.5$), as this region is dominated by photons from neutral meson decays; we call this correlation $C_\text{BR}$. To get the prompt photon correlation $C_\text{S}$, we do the following weighted subtraction with purity $P$:

\begin{equation}
    C_\text{S} = \frac{C_\text{SR} - (1-P)C_\text{BR}}{P}
    \label{eq:brsub}
\end{equation}

Four correlations are measured with the purity weight described in Eq.~\ref{eq:brsub} and combined to get the fully-subtracted correlation signal: the same-event signal-region (SESR), the same-event background-region (SEBR), the mixed-event signal-region (MESR), and the mixed-event background-region (MEBR). As mentioned above, the background-region (BR) correlations account for the non-prompt photon background, while the mixed-event (ME) correlations account for the combinatorial background. To avoid double-counting the combinatorial background associated with non-prompt photons, we add MEBR back to get the final signal yield:

\begin{equation}
    \gamma_\text{prompt} + \text{correlated jets} = \text{SESR} - \text{SEBR} - \text{MESR} + \text{MEBR}
\end{equation}

The dominant sources of systematic uncertainty are the uncertainty associated with the shape of the mixed-event correlation and the uncertainty associated with the selection of the background-region used to estimate the non-prompt photon background. To evaluate the mixed-event shape uncertainty, we consider the difference in the shape of the $\deltaphi$ correlations between the same-event and mixed-event correlations in the region with minimal signal. To evaluate the background-region uncertainty, we vary the $\ssvar$ range used to measure $C_\text{BR}$.

\section{Results}
\begin{figure}[htb]
\centerline{%
\includegraphics[width=12.5cm]{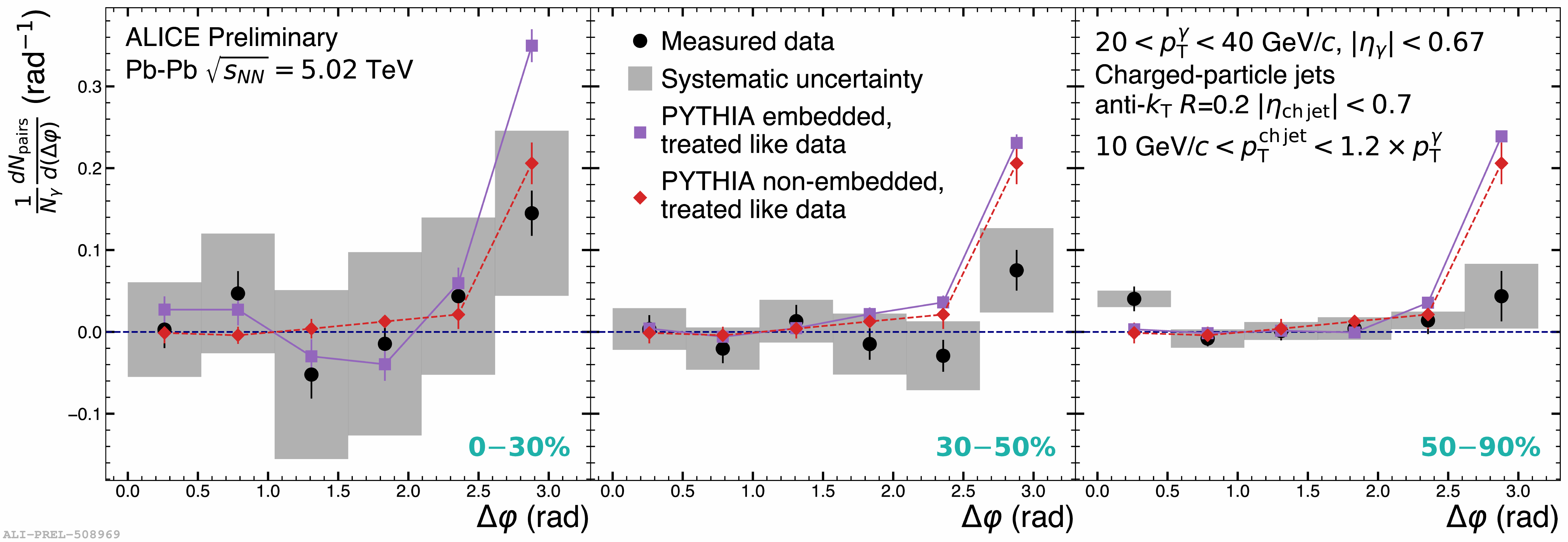}}
\caption{The angular correlations between isolated photons and jets are compared with embedded gamma-jet and dijet PYTHIA treated like data and also non-embedded gamma-jet and dijet PYTHIA treated like data}
\label{fig:deltaphi}
\end{figure}

The fully-subtracted angular correlations $\deltaphi$ are shown in \figurename~\ref{fig:deltaphi}. They are compared to PYTHIA embedded into minimum-bias Pb-Pb data, to account for the effects of the underlying event present in the Pb-Pb environment, as the data are not unfolded. They are also compared to a non-embedded PYTHIA simulation as an approximation of a pp reference. We see a trend of a higher back-to-back yield in more central collisions in both the data and the embedded PYTHIA, suggesting that this arises from underlying event fluctuations. We also see an apparent suppression in the most peripheral bin, but we note that 50\% is not very peripheral and there is much less ``extra'' yield from underlying event fluctuations, so this is not as surprising as it might seem.

\begin{figure}[htb]
\centerline{%
\includegraphics[width=12.5cm]{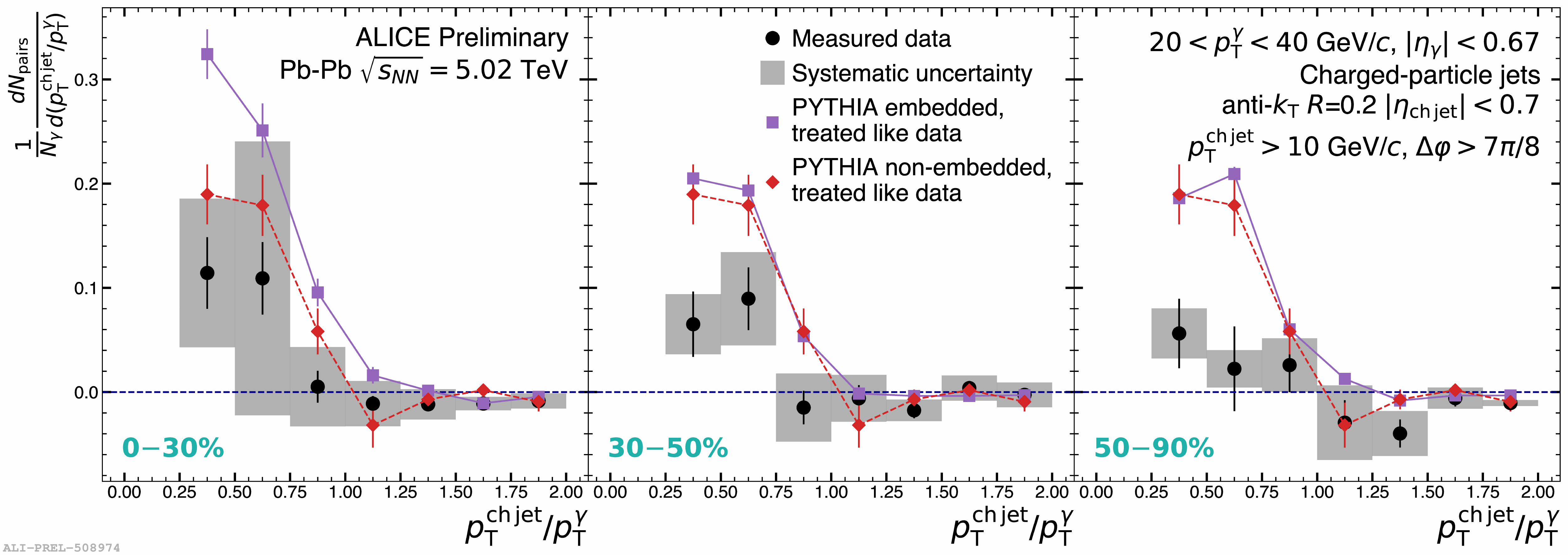}}
\caption{The momentum imbalance between isolated photons and jets is compared with embedded gamma-jet and dijet PYTHIA treated like data and also non-embedded gamma-jet and dijet PYTHIA treated like data}
\label{fig:ptratio}
\end{figure}

The fully-subtracted momentum imbalance correlations $\ptratio$ are shown in \figurename~\ref{fig:ptratio}, and, as with the $\deltaphi$ observable, are compared to both embedded and non-embedded PYTHIA. A shape difference as a function of centrality is seen in the embedded PYTHIA due to detector effects.

\begin{figure}[htb]
\centerline{%
\includegraphics[width=5.6cm]{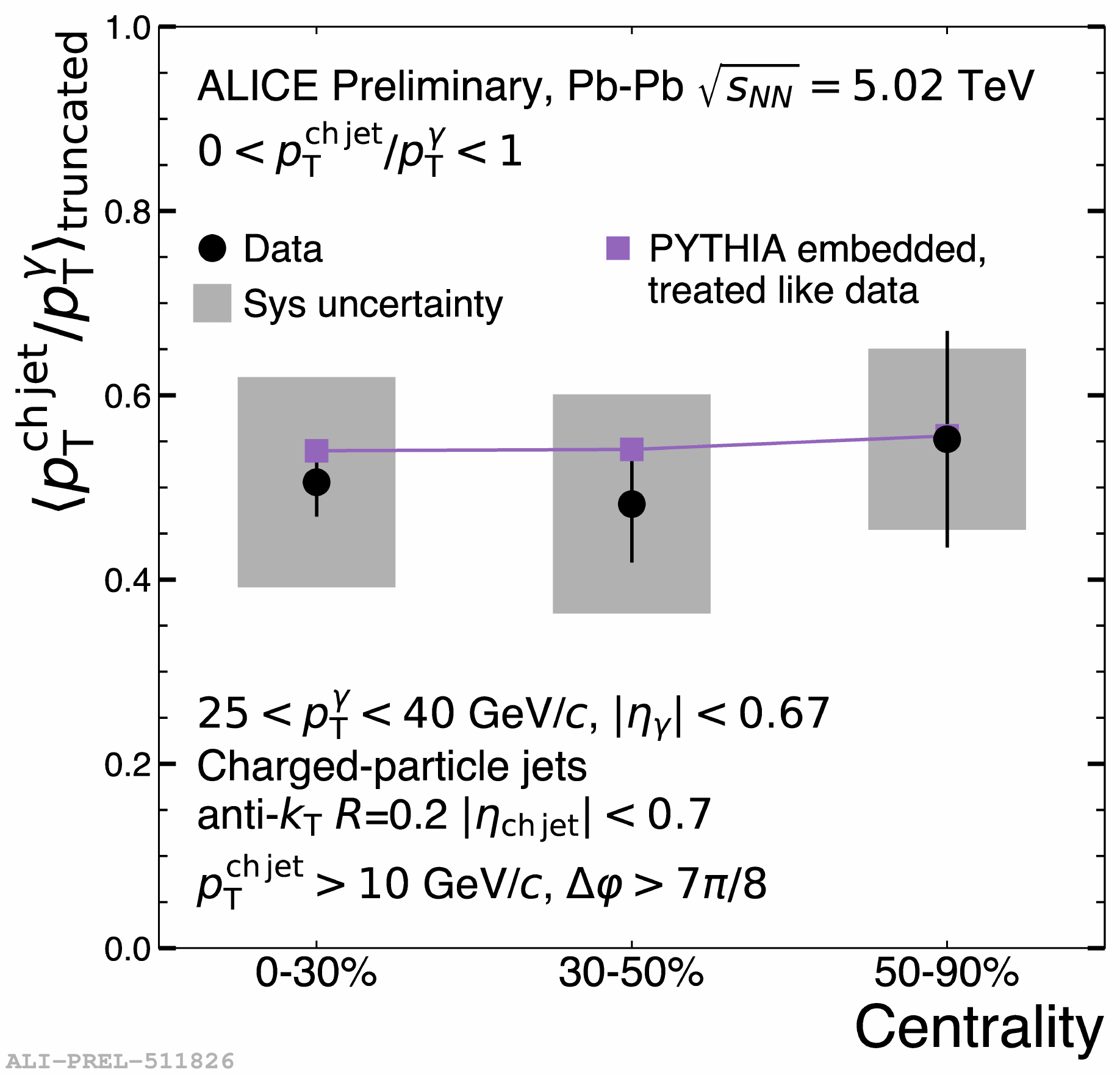}}
\caption{The truncated mean (between 0 and 1) of momentum imbalance between isolated photons and jets is compared with a combined embedded gamma-jet and dijet PYTHIA sample which has been treated like data.}
\label{fig:ptratiomean}
\end{figure}

In order to study the centrality dependence of jet energy loss, we calculate the mean of the $\ptratio$ distribution within $0 < \ptratio < 1$. This is shown in \figurename~\ref{fig:ptratiomean}. Within our uncertainties, we do not observe a centrality dependence. For higher-energy photons and jets, CMS and ATLAS reported modifications for more central Pb-Pb collisions compared to a pp reference.

\section{Summary}
We present the first measurement of isolated photon-jet correlations in Pb-Pb collisions in ALICE and the first such measurement at the LHC down to photon $\pt = 20$ \GeVc. This extends the LHC analyses down to a lower $Q^2$ and $x$ regime. No centrality-dependent medium modification is seen for the photon-jet angular correlation or $\pt$ imbalance within the uncertainties. We plan to compare to model predictions such as CoLBT and are looking forward to more precise measurements with the upcoming LHC Run 3 data.


\printbibliography

@article{gammahadron,
  title = {Measurement of isolated photon-hadron correlations in $\sqrt{{s}_{\mathrm{NN}}}$ = 5.02 {TeV} $pp$ and $p$-{Pb} collisions},
  author = {ALICE Collaboration},
  journal = {Phys. Rev. C},
  volume = {102},
  issue = {4},
  pages = {044908},
  numpages = {16},
  year = {2020},
  month = {Oct},
  publisher = {American Physical Society},
}

@article{gammajet-cms,
title = {Study of jet quenching with isolated-photon+jet correlations in {PbPb} and pp collisions at {sNN=5.02 TeV}},
journal = {Physics Letters B},
volume = {785},
pages = {14-39},
year = {2018},
issn = {0370-2693},
doi = {https://doi.org/10.1016/j.physletb.2018.07.061},
url = {https://www.sciencedirect.com/science/article/pii/S0370269318306245},
author = {CMS Collaboration}
}

@article{gammajet-atlas,
title = {Measurement of photon–jet transverse momentum correlations in 5.02 {TeV} {Pb + Pb} and pp collisions with {ATLAS}},
journal = {Physics Letters B},
volume = {789},
pages = {167-190},
year = {2019},
issn = {0370-2693},
doi = {https://doi.org/10.1016/j.physletb.2018.12.023},
url = {https://www.sciencedirect.com/science/article/pii/S037026931830950X},
author = {ATLAS Collaboration}
}

@article{fastjetmanual,
   title={{FastJet} user manual},
   volume={72},
   ISSN={1434-6052},
   url={http://dx.doi.org/10.1140/epjc/s10052-012-1896-2},
   DOI={10.1140/epjc/s10052-012-1896-2},
   number={3},
   journal={The European Physical Journal C},
   publisher={Springer Science and Business Media LLC},
   author={Cacciari, Matteo and Salam, Gavin P. and Soyez, Gregory},
   year={2012},
   month={Mar} }

@article{jetareamedian,
    author = "Cacciari, Matteo and Salam, Gavin P.",
    title = "{Pileup subtraction using jet areas}",
    eprint = "0707.1378",
    archivePrefix = "arXiv",
    primaryClass = "hep-ph",
    reportNumber = "LPTHE-07-01",
    doi = "10.1016/j.physletb.2007.09.077",
    journal = "Phys. Lett. B",
    volume = "659",
    pages = "119--126",
    year = "2008"
}

@article{alice-experiment,
	doi = {10.1088/1748-0221/3/08/s08002},
	url = {https://doi.org/10.1088/1748-0221/3/08/s08002},
	year = 2008,
	month = {aug},
	publisher = {{IOP} Publishing},
	volume = {3},
	number = {08},
	pages = {S08002--S08002},
	author = {ALICE Collaboration},
	title = {The {ALICE} experiment at the {CERN} {LHC}},
	journal = {Journal of Instrumentation}
}

@article{alice-performance,
author = {ALICE Collaboration},
title = {Performance of the {ALICE} experiment at the {CERN} {LHC}},
journal = {International Journal of Modern Physics A},
volume = {29},
number = {24},
pages = {1430044},
year = {2014},
doi = {10.1142/S0217751X14300440},

URL = { 
        https://doi.org/10.1142/S0217751X14300440
    
},
eprint = { 
        https://doi.org/10.1142/S0217751X14300440
    
}
}

\end{document}